# High-Bandwidth and Large Coupling Tolerance Graded-Index Multimode Polymer Waveguides for On-board High-Speed Optical Interconnects

Jian Chen, *Student Member, IEEE*, Nikolaos Bamiedakis, Peter P. Vasil'ev, Tom J. Edwards, Christian T.A. Brown, Richard V. Penty, *Senior Member, IEEE*, and Ian H. White, *Fellow, IEEE*

*Abstract*—Optical interconnects have attracted significant research interest for use in short-reach board-level optical communication links in supercomputers and data centres. Multimode polymer waveguides in particular constitute an attractive technology for on-board optical interconnects as they provide high bandwidth, offer relaxed alignment tolerances, and can be cost-effectively integrated onto standard printed circuit boards (PCBs). However, the continuing improvements in bandwidth performance of optical sources make it important to investigate approaches to develop high bandwidth polymer waveguides. In this paper, we present dispersion studies on a graded-index (GI) waveguide in siloxane materials designed to deliver high bandwidth over a range of launch conditions. Bandwidth-length products of >70 GHz×m and ~65 GHz×m are observed using a 50/125 μm multimode fibre (MMF) launch for input offsets of ±10 μm without and with the use of a mode mixer respectively; and enhanced values of >100 GHz×m are found under a 10× microscope objective launch for input offsets of ~18 × 20 μm². The large range of offsets is within the -1 dB alignment tolerances. A theoretical model is developed using the measured refractive index profile of the waveguide, and general agreement is found with experimental bandwidth measurements. The reported results clearly demonstrate the potential of this technology for use in high-speed board optical links, and indicate that data transmission of 100 Gb/s over a multimode polymer waveguide is feasible with appropriate refractive index engineering.

## I. INTRODUCTION

CONVENTIONAL copper-based electrical links must overcome challenges to meet the need for ever increasing interconnection bandwidth in next generation high-performance electronic systems such as supercomputers and data centres [1]. Electrical interconnects have inherent limitations when operating at high frequencies such as high losses, limited bandwidth, poor crosstalk and high power consumption [2]–[5]. Optical technologies enable larger bandwidth, immunity to electromagnetic interference (EMI) and lower power consumption [6]. As a result, optical interconnects have attracted considerable research for use in board-level communication links in recent years. For instance, 25 Gb/s optical interconnects are becoming commercially available, while 40 Gb/s on-off keying (OOK)-based and 56 Gb/s 4-level pulse amplitude modulation (PAM-4)-based optical interconnections have also been demonstrated [7]–[9]. Several optical technologies have been proposed for short-reach communication links, including fibre technologies [10]–[12], free-space optics [13]–[15], Si photonics [16], [17] and polymer waveguides [18], [19].

Polymer waveguides are promising candidates for use in board-level interconnection links as they can be directly integrated onto printed circuit boards (PCBs) using conventional methods employed by the electronics industry [20]. The polymer materials exhibit favourable mechanical and thermal properties to withstand fabrication of standard PCBs. The use of multimode waveguides with their large size (30 × 30 μm² to 70 × 70 μm²) offers relaxed alignment tolerances and allows system assembly with common pick-and-place tools (typically with alignment tolerances of better than ±10 μm) [21].

Vertical-cavity surface-emitting lasers (VCSELs) are typically employed in conjunction with these multimode polymer waveguides as they are cost-effective, power-efficient, exhibit high bandwidth, and can be formed in large arrays [22]. The bandwidth performance of VCSELs has been continuously improving over the past few years with most recent results demonstrating error-free transmission up to 57 Gb/s without the use of equalization, and 71 Gb/s when equalization schemes are applied [23]–[27]. Important concerns have been raised on the potential of the above polymer waveguides to support very high data rates and keep up with the improvements in the bandwidth of VCSELs. In addition, the requirement for relaxed alignment tolerances in the production and assembly of the interconnection system makes it imperative to assert that the required channel bandwidth is available for a large range of input positions and under various types of launches.

Graded-index polymer waveguides have therefore been proposed to achieve higher bandwidth compared with step-index multimode waveguides [28], [29]. In this paper therefore, we demonstrate a graded-index multimode polymer waveguide using siloxane materials and its bandwidth limit is

J. Chen, N. Bamiedakis, Peter P. Vasil'ev, R. V. Penty and I. H. White, are with the Electrical Engineering Division, Engineering Department, University of Cambridge, CB3 0FA, Cambridge, UK (tel: 0044-1223748363; e-mail: jc791@cam.ac.uk).

Tom J. Edwards, Christian T.A. Brown are with SUPA, School of Physics & Astronomy, University of St Andrews, North Haugh, St Andrews, Fife, KY16 9SS, UK.

explored theoretically and experimentally under various launch conditions with spatial input offsets at two wavelengths.

Dispersion studies on multimode polymer waveguides have been carried out by different groups in the past. For example, the estimated -3 dB bandwidth and bandwidth-length product (BLP) of the guides reported has been found to be 23 GHz (BLP: 57.5 GHz×m) for a 2.55 m long waveguide [30] and 150 GHz (BLP: 75 GHz×m) for a 51 cm long waveguide [31] under a single-mode fibre (SMF) launch. A larger value of 1.03 GHz (BLP: 90 GHz×m) for a 90 m long graded-index waveguide have been found under restricted launch [32]. These measurements however, have been carried out only for a centre launch under restricted launch condition, while the effect of launch conditions and input offsets on bandwidth performance has not been reported. We have previously presented bandwidth studies on a 1 m long spiral multimode polymer waveguide using frequency-domain ($S_{21}$) measurements and reported a bandwidth-length product (BLP) larger than the measurement limit of 35 GHz×m even in the presence of input spatial offsets and for various launch conditions [33]. These measurements were limited by the capability of the instruments and the bandwidth of the active devices used in the experiments. As a result, we have undertaken time-domain (pulse broadening) measurements in order to measure the actual bandwidth of these guides and assess their ability to support data rates above 100 Gb/s. In the work presented herein, we report experimental and simulation studies on the dispersion of new graded-index waveguides including a 105.5 cm long multimode polymer spiral waveguide and a 19.2 cm long reference waveguide. The obtained results demonstrate BLPs in excess of 100 GHz×m under restricted launch and BLPs in excess of 70 GHz×m even for a 50/125 µm multimode-fibre (MMF) launch in a large range of input offsets.

In addition, the coupling loss performance of these guides is assessed for the launch conditions studied and large -1 dB alignment tolerances are demonstrated ($\geq \pm 10$ µm). The range of input offsets that provides both high bandwidth and low coupling losses is identified and it is found to be ~18 × 20 µm². This result highlights the strong potential of these guides for use in real-world systems.

The paper is structured as follows. Section II introduces the multimode polymer waveguides employed in this work while Section III and IV present the experimental and theoretical bandwidth studies respectively. Section V includes a brief discussion on the reported results while Section VI draws the conclusions.

## II. MULTIMODE POLYMER WAVEGUIDES

The waveguide samples employed in this work are fabricated from siloxane materials developed by Dow Corning® on 8-inch silicon substrates using standard photolithography and comprise long spiral and shorter reference waveguides. The core and cladding materials are Dow Corning® WG-1020 Optical Waveguide Core and XX-1023 Optical Waveguide Clad respectively. These polymer materials can be directly integrated on PCBs as they can withstand the high temperature (in excess of 250°C) required for solder reflow and board lamination. The particular materials used in this work have been optimised for low-loss operation at short data communications wavelengths (0.8 − 1 µm). For example, the loss is ~0.04 dB/cm at 0.85 µm and ~0.1 dB/cm at 1 µm [34]. The attenuation shows no measurable change after 2000 h at 85 °C and in a 85% relative humidity environment, or after 500 thermal cycles from -40 °C to +120 °C [35], [36]. The spiral waveguides have a length of 105.5 cm and the reference waveguides have a length of 19.2 cm. The cross section of the waveguides is approximately 35 µm × 35 µm. The waveguide samples are fabricated to exhibit a graded-like refractive index profile by adjusting fabrication parameters [34], [37]. Fig. 1(a) and Fig. 1(b) show photographs of the spiral and the reference waveguide employed in this work illuminated with a red laser source.

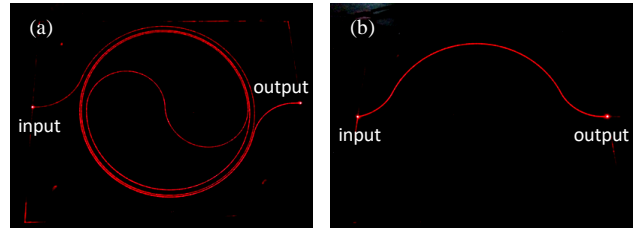

Fig. 1 Photograph of (a) the 105.5 cm long spiral and (b) the 19.2 cm long reference waveguide illuminated with red light.

The refractive index profile of the waveguide samples is measured using the refractive near field method and is shown in Fig. 2 [38]. It exhibits a triangle-like shape toward the upper side of the core and the maximum index difference between the core and cladding is $\Delta n \sim 0.01$. The index variation of the cladding seen in the plot [Fig. 2(a)] is due to measurement artifacts.

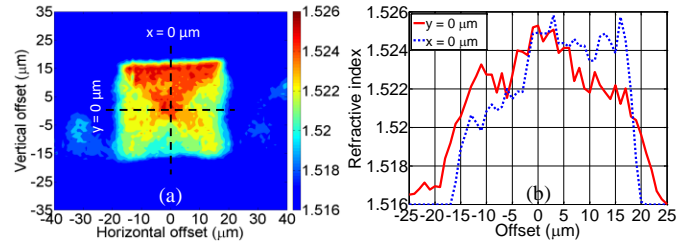

Fig. 2 Measured refractive index profile of the GI waveguides at the wavelength of 678 nm: (a) 2-dimensional; (b) 1-dimensional when vertical offset y = 0 µm (red solid) and horizontal offset x = 0 µm (blue dashed).

The refractive index profile of the waveguides, described as "graded-index" (GI) in the rest of this paper although it does not have the parabolic GI profile typically encountered in optical fibres. The index grading is found to vary with fabrication conditions and material formulations. Detailed studies carried out by Dow Corning have determined that this graded index profile can be reliably produced with low variability. The particular profile used in these waveguides are designed for low coupling loss with 50/125 µm MMF inputs/outputs and low crossing loss, while the studies presented here demonstrate that very good bandwidth performance is also achieved. The improved performance over waveguides with a step-like profile is due to the lower number of modes guided and therefore the reduced levels of induced multimode dispersion.

## III. EXPERIMENTAL STUDIES

### A. Experimental setups

To allow more detailed study with a range of launch conditions, two short pulse generation systems are employed in the experiments in order to measure the dispersion induced by the waveguides. The first system utilises a femtosecond Ti:Sapphire laser operating at a wavelength of 850 nm as the source, and a FR103-MN autocorrelator as the detector. The second one employs a femtosecond erbium-doped fibre laser source (TOPTICA FemtoFiber Scientific) operating at a wavelength of $1574\pm0.5$ nm and a frequency-doubling crystal (MSHG1550-0.5-1) to generate short pulses at the wavelength of $787\pm0.5$ nm. A matching autocorrelator is used to detect the transmitted signals at the receiver end. Fig. 3(a-c) illustrate the experimental setups used in the time-domain measurements carried out on the link with and without (back-to-back) the waveguide using the two short-pulse generation systems [39], [40].

The dispersion in the waveguide depends on the launch condition at the waveguide input, as this affects the generated mode power distribution inside the waveguide and therefore, the level of induced intermodal dispersion. Different spatial offsets are introduced as different launch positions excite different mode groups in the waveguide, thus affecting the induced dispersion. The measurements on the spiral waveguides are conducted for two launch conditions using a $50/125$ μm GI MMF: (a) without and (b) with a mode mixer (MM: Newport FM-1). The measurement on the reference waveguides is conducted using a $10\times$ [numerical aperture (NA)=0.25] microscope objective launch. For launch conditions (a) and (b), the light is coupled to a $50/125$ μm multimode fibre (MMF) patchcord (<1 m) via a $10\times$ (NA=0.25) microscope objective. Fig. 3(d-f) show the near field images of the input for each launch condition studied. As it can be observed, there is only a small number of mode groups excited inside the input MMF patchcord under a centre lens launch for launch condition (a) [Fig. 3(d)]. For launch condition (b), the use of the MM results in the distribution of a greater proportion of power to higher-order modes inside the fibre and hence, produces a "worst-case" launch condition with respect to dispersion [Fig. 3(e)]. The restricted launch results in a Gaussian input with a spot size of $5\pm1$ μm [full-width-at-half-maximum (FWHM)] [Fig. 3(f)].

The cleaved end of the input MMF or the $10\times$ microscope objective is mounted on a precision translation stage, while the introduced spatial offsets are controlled via a precision displacement sensor. A $16\times$ microscope objective (NA=0.32) is used to collect the light at the waveguide output and feed it to the autocorrelator. The $16\times$ lens is chosen as its NA is larger than that of the waveguide preventing thus, any mode selective loss at the waveguide output. The optical power received at the waveguide output is measured using a broad area power optical head (HP 81525A) while the autocorrelation trace of the transmitted pulses is recorded for each launch condition studied and for the different input positions.

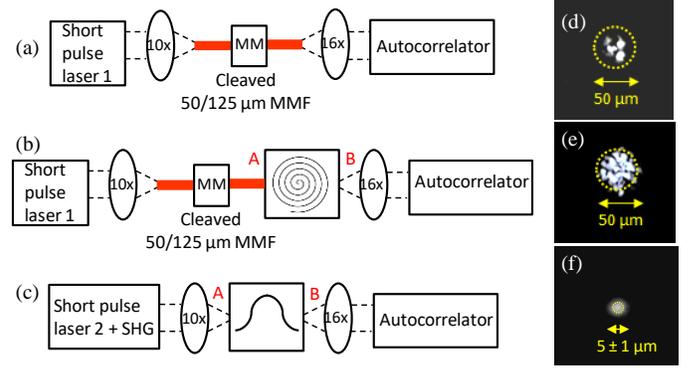

Fig. 3  Experimental setup used for the (a) back-to-back link; (b) 105.5 cm long spiral waveguide link under a $50/125$ μm MMF launch with the use of a mode mixer (MM) using pulse generation system 1 at λ = 850 nm; (c) 19.2 cm long reference waveguide link under a $10\times$ microscope objective launch using pulse generation system 2 at λ = 787 nm; and near field images of input fibre and lens under the different launch conditions studied: (d) a $50/125$ μm GI MMF; (e) a $50/125$ μm GI MMF with a MM; (f) a $10\times$ microscope objective.

### B. Experimental results and discussion

The pulse width at the waveguide output is estimated from the recorded autocorrelation traces. The shape of the output pulses for each measurement is approximated with common pulse shapes (i.e. sech², Gaussian or Lorentzian) using curve fitting. The best-matching shape for each pulse is determined by choosing the one with minimum root-mean-square-error (RMSE) compared with the actual pulse. It should be noted that, although the output pulses of the femtosecond laser used in the experiments have a sech² shape, the output pulses after propagation over the multimode waveguides are strongly structured due to the waveguide dispersion. As a result, using the standard approach, we fit pulse shapes which have analytic solutions to the correlation equation to estimate the waveguide bandwidth. The received pulse is compared with that obtained for the respective back-to-back link in order to estimate the pulse broadening due to the light propagation through the waveguide. The frequency response of the back-to-back and the waveguide link for each launch condition and input position is calculated by taking the Fourier transform of the respective received optical pulses, and the waveguide frequency response of the waveguide is obtained by deconvolution. The -3 dB bandwidth of the waveguide can then be found for each particular input position and launch condition. Fig. 4 shows the -3 dB bandwidth of the 105.5 cm long spiral waveguide sample and the respective normalised received power as a function of the horizontal offset for a $50/125$ μm MMF launch with and without the use of the MM. The insertion loss of the GI spiral waveguide [power difference between the point A and B in Fig. 3(b)] is ~5.3 dB and ~6.6 dB under launch condition (a) and (b) respectively.

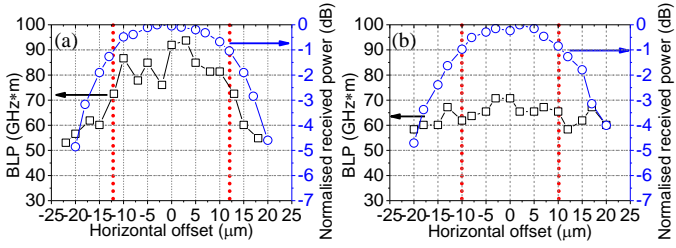

Fig. 4 Experimental results of the -3 dB bandwidth for the 105.5 cm long spiral polymer multimode waveguides and respective normalised received power for a 50/125 µm MMF launch with and without the MM: (a) GI waveguide, no MM (b) GI waveguide, MM (λ = 850 nm).

The BLP of the GI spiral waveguide is found to be >70 GHz×m using the 50/125 µm MMF launch for spatial offsets for a -1 dB alignment tolerances (~ ±10 µm) [Fig. 4(a)]. The waveguide bandwidth gradually reduces for larger input offsets, when a larger portion of the input power is coupled to higher-order modes at the waveguide input. A BLP of >65 GHz×m is obtained for the GI spiral waveguide across the offset range of ±10 µm for the mode-mixed 50/125 µm MMF launch. The use of the MM degrades the bandwidth performance due to the generation of a more uniform mode power distribution at the waveguide input [Fig. 3(e)], but also results in a more uniform dispersion performance across input offsets [Fig. 4(b)]. The induced spatial offsets have a small effect on the observed bandwidth as the generated mode power distribution does not vary significantly across input offsets due to the use of a relatively overfilled launch.

Similar measurements are carried out on the 19.2 cm long reference waveguides using a 10× microscope objective launch. The insertion loss of the reference waveguide [power difference between the point A and B in Fig. 3(c)] is ~1.5 dB under such a launch. This condition provides a restricted launch at the waveguide input as the input beam has a small spot-size and only excites a limited number of modes inside the waveguide. As a result, an improved bandwidth performance is expected. Fig. 5 shows the near-field images of the reference waveguide output for the lens launch [Fig. 5(a)] as well as the one obtained at the spiral output for a 50/125 µm MMF launch without the MM for comparison [Fig. 5(b)]. These images justify that 10× lens launch offers a highly selective mode excitation in the waveguide whereas a larger percentage of power is distributed towards the outer edge of the waveguide using a 50/125 µm MMF launch. A similar method as the one employed for the spiral waveguides is used to estimate the bandwidth of these reference waveguides for this restricted launch condition.

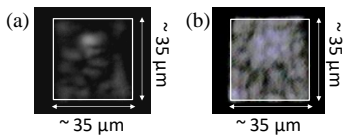

Fig. 5 Near field images at the output of (a) the 19.2 cm long reference GI waveguide under restricted launch; (b) the 105.5 cm long spiral GI waveguide under a 50/125 µm MMF launch without the MM.

Indeed, the obtained BLP for the reference 19.2 cm long GI waveguide is found to be >100 GHz×m for a large region of input offsets of ~18 × 20 µm² [Fig. 6(a)]. Fig. 6(b) shows that the high BLP region is also within the -1 dB alignment tolerances. The high-bandwidth area matches the higher refractive index of the waveguide at these launch positions (Fig. 2) and is located towards the upper side of the core. The large dimension of the area (>10 × 10 µm²) ensuring a bandwidth of >100 GHz×m and high coupling efficiency, indicates a robust bandwidth performance across a large range of input positions and demonstrates that achievable launch conditioning can be implemented in such highly-multimoded waveguides to ensure reliable operation. The use of restricted launches via launch conditioning schemes is a useful tool for improving bandwidth performance and ensuring reliable operation in multimode waveguide systems. It has been widely used in MMF systems to meet specifications for Gigabit Ethernet (GbE) and 10 GbE data transmission [41].

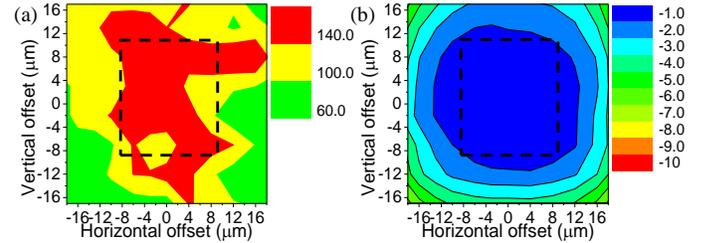

Fig. 6. (a) Experimental results of the -3 dB BLP (key in GHz×m units) (high BLP region noted in black dashed line); and (b) normalised measured coupling loss (units in dB) of the 19.2 cm long reference GI polymer multimode waveguide under a 10× microscope objective launch at λ = 787 nm (black dashed line indicates the high BLP region shown in (a)).

## IV. NUMERICAL STUDIES

This section describes numerical simulations carried out to understand the performance of the waveguides. For large polymer multimode waveguides, chromatic dispersion is much less than multimode dispersion [42]. Thus no chromatic dispersion is considered in the simulation modeling.

### A. Simulation method

The simulation is based on standard methods employed for assessing MMF waveguides [43] and is presented in more detail in Appendix [42]. The field profile of the guided waveguide modes as well as their corresponding group refractive indices are calculated with a commercially-available mode solver using the measured refractive index profile of the waveguides. The launch is assumed to consist of the modes of a 50/125 µm MMF which are obtained using an analytic mode solver assuming the standard refractive index profile of a MMF. The calculated number of modes for a standard 50/125 µm MMF is 184 with a maximum principle mode number (PMN) of 29. The exact mode power distribution inside the input MMF patchcord in the experiment is unknown and is hard to be accurately measured. Therefore, the power is assumed to be distributed equally among the modes of the MMF patchcord with a PMN≤15 for the launch without the MM, and with a PMN≤22 for the launch with the MM (Fig. 7). The restricted launch is modeled with a Gaussian beam with a spot size of the measured value (FWHM~4 µm). Overlap integrals are used to calculate the power coupled to each waveguide mode for a given launch. The impulse response and therefore the bandwidth of the waveguide can then be calculated using the mode group refractive indices and mode power. Mode selective loss is introduced in the model to account for the suppression of higher-order modes due to the waveguide bends of the samples. This is found by fitting the simulation with the experimental data of the received optical power across spatial offsets (Fig. 8). The process is repeated for the different launch conditions at different input positions and the respective waveguide bandwidth is calculated from the obtained impulse response for each offset under each launch condition.

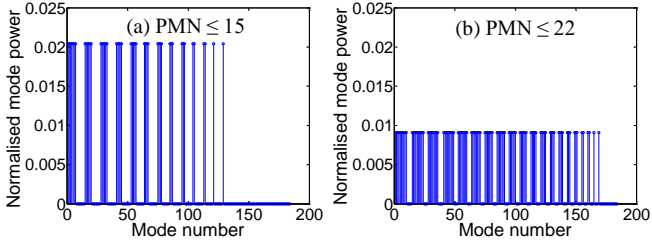

Fig. 7. Assumed mode power distribution inside the input 50/125 μm MMF patchcord for condition: (a) a 50/125 μm MMF with no MM (PMN ≤ 15); (b) a 50/125 μm MMF with MM (PMN ≤ 22).

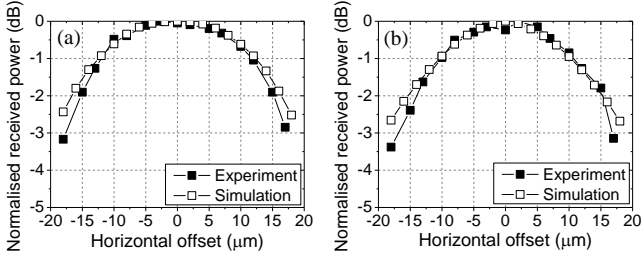

Fig. 8 Normalised received power versus position of input 50/125 μm MMF: (a) a 50/125 μm MMF with no MM (PMN ≤ 15); (b) a 50/125 μm MMF with MM (PMN ≤ 22).

### B. Simulation results

The obtained simulation results for the spiral waveguides under the MMF launch are shown in Fig. 9. It can be seen that a similar behaviour is obtained as the one observed in the experimental work. Fig. 10 shows the simulation result for the reference waveguide under restricted launch. The high BLP region obtained from the simulation is in general agreement with the experimental observations.

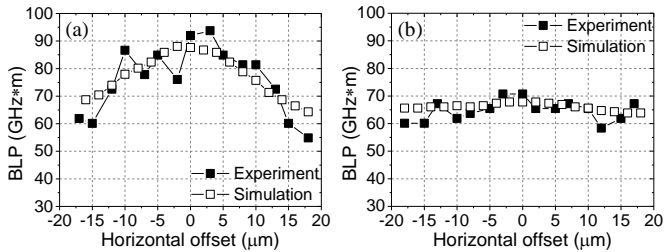

Fig. 9. Simulation and experimental results for the 105.5 cm long spiral polymer multimode waveguides for a 50/125 μm MMF input with and without MM: (a) GI waveguide, no MM; (b) GI waveguide, MM (λ = 850 nm).

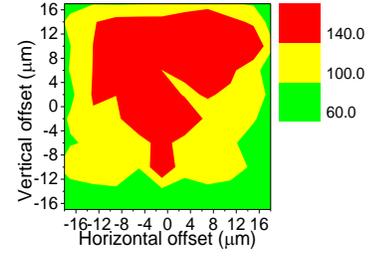

Fig. 10. Simulation results of the -3 dB BLP (unit in GHz×m) of the 19.2 cm long reference GI polymer multimode waveguide for a 10× microscope objective launch at λ = 787 nm.

## V. DISCUSSION

### A. Simulation vs. experimental results

For the particular studies presented here, the successful fitting of the measured bandwidth performance indicates that the simulation studies can provide a guideline for launch design and dispersion engineering. In the case of multimode launches, such as the ones reported above, however, it is particularly difficult to know in practice the exact mode power distribution at the waveguide input and therefore calculate the induced dispersion unless specific launch conditioning schemes are employed. Nevertheless, as performed in the study of Gigabit Ethernet (GbE) and 10 GbE standards in MMF links [41], such a model can provide a good indication of the expected waveguide performance. It should be noted however, that there are a number of phenomena that affect the mode propagation inside the waveguide which have not been taken into account in the simulations: (a) the calculated mode profiles are based on an ideal straight waveguide whereas the actual mode profiles inside the complex long bending waveguide samples may differ; (b) mode coupling is not considered. It has been experimentally verified previously in different polymer material systems that multimode dispersion in such guides can be significantly affected by mode mixing with very short equilibrium lengths of ~10 cm observed [30]. In our case however, the strong dependence of induced dispersion on launch conditions observed in the experimental results, suggests that the waveguide lengths studied (up to 1 m) are shorter than the equilibrium length of these particular waveguides. Further studies are required to assess the magnitude of these phenomena and their effect on the mode propagation inside the waveguides. In any case, these experimental and simulation results strongly demonstrate these particular waveguides exhibit an excellent bandwidth performance. Moreover, one may further improve the bandwidth performance of the polymer waveguides by optimising the refractive index distribution of the core.

### B. Towards 100 Gb/s on-board links

For the successful deployment of NRZ-based 100 Gb/s on-board links, the multimode waveguides not only need to exhibit adequate bandwidth but also sufficiently low insertion loss to ensure that adequate power reaches the receiver. A calculation using some basic assumptions is carried out to provide an indication of the necessary power budgets. A moderate launch power of 6 dBm is assumed for the laser source while the 60 GHz receiver sensitivity is calculated to be -3 dBm using the same noise spectral density as a commercial 30 GHz multimode receiver (Newport 1474-A, noise-equivalent-power of 38 pW/$\sqrt{\text{Hz}}$), yielding a power budget of 9 dB. Assuming a loss of 0.04 dB/cm for the waveguide, for a 1 m long link, there is an adequate 5 dB of power margin for coupling and component losses.

## VI. Conclusion

Multimode polymer waveguides constitute a promising technology for on-board optical interconnects as they can be directly implemented on PCBs and offer cost-effective assembly solutions. The continuous improvement of VCSELs has raised the questions on the capacities of these waveguides due to their highly-multimoded nature. In this paper, we have conducted time-domain measurements to investigate the ultimate bandwidth of GI waveguides and shown bandwidth-length products of >70 GHz×m and ~65 GHz×m using a 50/125 μm MMF launch for an offset range of ±10 μm without and with the use of a mode mixer respectively, and of >100 GHz×m using a restricted launch for an offset range of ~18 × 20 μm². Moreover, the high bandwidth regions are well within -1 dB alignment tolerances. A model based on the measured refractive index profile is developed, and the simulation results show general agreement with the experimental results. These results indicate the potential of this technology and the capacity of transmitting data rates of 100 Gb/s over a single waveguide channel up to distances of 1 m with appropriate engineering of refractive index.

## Acknowledgment

The authors would like to acknowledge Dow Corning for providing the waveguide samples and EPSRC via the Complex Photonic Systems II (COPOS II) project for supporting the work. Additional data related to this publication is available at the University of Cambridge data repository (https://www.repository.cam.ac.uk/handle/1810/248987).

## Appendix

This section describes the calculation procedure used in the simulation. The effective refractive indices $n_{eff}^i(\lambda_0)$ of all the guided waveguide modes are computed using commercial software package FIMMWAVE based on the measured refractive index profiles as shown in Fig. 2, and the group refractive indices $n_{group}^i(\lambda_0)$ of the modes are then calculated using the expression:

$$n_{group}^i(\lambda_0) = n_{eff}^i(\lambda_0) - \lambda_0 \cdot \frac{dn_{eff}^i}{d\lambda}\Big|_{\lambda = \lambda_0} \qquad (1)$$

The respective propagation delay $t_i$ of each waveguide mode for a given link length L can be found:

$$t_i = \frac{L \cdot n_{group}^i}{c} \qquad (2)$$

where c is the speed of light in vacuum.

Thus, the impulse response h(t) of the waveguide can be expressed as:

$$h(t) = \sum_i p_i a_L \cdot \delta(t - t_i) \qquad (3)$$

where $p_i$ is the mode power distribution (i = 1,…, N where N is the total number of guided modes), $a_L$ is the power attenuation of the modes. The mode power distribution inside the waveguide can be calculated using the overlap integral of the electric fields of the input fibre modes and the waveguide modes [43]. The mode power attenuation ($a_L$) of the waveguide sample is chosen to be a step function as an approximation to eliminate the higher-order modes beyond a certain mode number. It is found by fitting the measured received power as a function of input offsets to the calculated coupling loss of a straight waveguide.

The electrical field distributions of the guided fibre modes inside the MMF and the corresponding propagation constants are calculated using an analytic mode solver. The refractive index profile of the MMF and the operating wavelength are the input parameters of the calculation, and subsequently the linear polarized (LP) modes with indices m (cylindrical order) and n (radial order) can be obtained [44]. The modes can be divided into degenerated groups indexed by the PMN defined as m + 2n + 1. The power coupling coefficient $c_{ij}$ (i.e. power coupled into each waveguide mode $i$ from each fibre mode $j$) can be expressed as (in polar coordinates):

$$c_{ij} = \frac{\left|\iint E_i^w E_j^f \, rdrd\phi\right|^2}{\left(\iint \left|E_i^w\right|^2 rdrd\phi\right) \cdot \left(\iint \left|E_j^f\right|^2 rdrd\phi\right)} \qquad (4)$$

where $E_i^w$, $E_j^f$ are the electric fields of the waveguide mode $i$ and fibre mode $j$ respectively.

The power $p_i^w$ coupled into each waveguide mode i can be expressed as the multiplication of the coefficient matrix $c_{ij}$ and the power $p_j^f$ of the fibre modes ($j = 1,…, M$ where M is the number of fibre modes):

$$p_i^w = \sum_{j=1}^M c_{ij} p_j^f \qquad (5)$$

where $p_j^f$ =1/M if assuming a uniform power distribution inside the input fibre.

The frequency response H(f) of the waveguide can be calculated using a Fourier transform and the -3 dB bandwidth can be derived from the normalised frequency response.